\documentclass[twocolumn,superscriptaddress,showpacs,prl]{revtex4}
\usepackage{mathrsfs}
\usepackage{amssymb, amsbsy, amsmath, latexsym, dsfont, array, layout, graphicx,mathrsfs,color}

\newcommand{\one}{\mathds{1}}
\newcommand{\ket}[1]{\left|{#1}\right\rangle}
\newcommand{\bra}[1]{\left\langle{#1}\right|}

\begin{document}

\title{Experimental realization of a generalized measuring device via a one-dimensional photonic quantum walk}
\author{Zhihao Bian}
\affiliation{Department of Physics, Southeast University, Nanjing
211189, China}
\author{Rong Zhang}
\affiliation{Department of Physics, Southeast University, Nanjing
211189, China}
\author{Hao Qin}
\affiliation{Department of Physics, Southeast University, Nanjing
211189, China}
\author{Xiang Zhan}
\affiliation{Department of Physics, Southeast University, Nanjing
211189, China}
\author{Jian Li}
\affiliation{Department of Physics, Southeast University, Nanjing
211189, China}
\author{Peng Xue\footnote{gnep.eux@gmail.com}
} \affiliation{Department of Physics, Southeast University, Nanjing
211189, China} \affiliation{State Key Laboratory of Precision
Spectroscopy, East China Normal University, Shanghai 200062, China}

\begin{abstract}
We demonstrate an implementation of unambiguous state discrimination
of two equally probable single-qubit states via a one-dimensional
photonic quantum walk. Furthermore we experimentally realize a
quantum walk algorithm for implementing a generalized measurement in
terms of positive operator value measurement on a single qubit. The
measurement of the single-photons' positions corresponds to a
measurement of an element of the positive operator value measurement
on the polarizations of the single-photons.
\end{abstract}

\maketitle

Quantum walks (QWs) exhibit distinct features compared to classical
random walks (RWs)~\cite{ADZ93} and hence can be used to develop
quantum algorithms~\cite{Ambainis,Spielman,Whaley02,Kempe03} and to
server as an ideal test-bed for studying quantum effects such as
Anderson
localization~\cite{W12,ES11,YKE08,K10,SK10,C13,SS11,ZXT14,ZX14,XQT14},
quantum chaos~\cite{WM04,BB04,B06,GA+13,XQTS14} and energy transport
in photosynthesis~\cite{OPR06,HSW10}. Furthermore, discrete-time QW
is a process in which the evolution of a quantum particle on a
lattice depends on a state of a coin. The coin degrees of freedom
offer the potential for a wider range of controls over the evolution
of the walker than are available in the continuous-time QW. Thus
there are more applications for discrete-time QW such as a versatile
platform for the exploration of a wide range of nontrivial
topological effects~\cite{KRBE10,A12,KB+13} which have been
implemented both theoretically and experimentally. Also QWs can be
considered as a generalized measurement proposed theoretically
in~\cite{KW13}. In such a scenario the measurement of the quantum
walker at a certain position $x = i$ corresponds to a measurement of
an element $E_i$ of a positive operator value measure (POVM) on a
coin state.


Quantum mechanics forbids deterministic discrimination among
nonorthogonal states. Nonetheless, the capability to distinguish
nonorthogonal states unambiguously is an important primitive in
quantum information processing. Unambiguous state discrimination
between $N$ states has $N+1$ outcomes: the $N$ possible conclusive
results, and the inconclusive result. Since no projective
measurement in an $N$-dimensional Hilbert space can have more than
$N$ outcomes, generalized measurements such as POVMs are required.
POVMs can be implemented by embedding the system into a larger
Hilbert space and unitarily entangling it with the extra degrees of
freedom (ancilla). Projective measurement of the ancilla induces an
effective non-unitary transformation of the original system. By an
appropriate design of the entangling unitary, this effective
non-unitary transformation can turn an initially nonorthogonal set
of states into a set of orthogonal states with a finite probability
of success. Based on this, the coupling between the walker and coin
can be used to implement the generalized measurement. The coin is
regarded as the target system whose states need to be discriminated.
The walker is regarded as the ancilla and the outcomes of the
projective measurement on it gives conclusive and inconclusive
results of the state discrimination. The POVM element is obtained by
taking the overlap between the final state of a properly engineered
QW and the initial state of the walker, and then tracing the walker
state out.

Unambiguous discrimination between two pure states has been
demonstrated in optical systems~\cite{HM96,CC01,MSB04,MCWB06,BF13}.
In this work, we realize an unambiguous state discrimination of two
equally probable single-qubit states via a one-dimensional
discrete-time QW, and implement a method suggested in~\cite{KW13}.
The QW indeed generates the POVM elements corresponding to the
unambiguous state discrimination problem.

The motivation of the paper is to use a very simple experiment to
show the meaning of the idea on using QW to implement a generalized
measurement which is shown in~\cite{KW13}. A QW is based on
projection measurement of the coin state, which can be extended in
the evolution time. The projection measurement model can naturally
be extend to an arbitrary generalized measurement. Compared to the
standard approach to POVM in which projection measurements are
performed on an extended Hilbert space, the extension of the Hilbert
space is not needed for the QW scenario.

First we explain how we realize an unambiguous state discrimination
of two pure single-qubit states through a QW optical-interferometer
network. The goal is to find one of the two single-qubit states
$\{\ket{0},\alpha\ket{0}+\beta\ket{1}\}$ with equal a priori
probabilities from three outcomes: two possible conclusive results,
and one inconclusive result.

The two states can always be represented by the two orthogonal
states as
\begin{equation}
\ket{\psi_\pm}=\cos \frac{\phi}{2}\ket{H}\pm\sin
\frac{\phi}{2}\ket{V},
\end{equation}
where $\phi\in\left[0,\pi/2\right]$, $\ket{H}$ ($\ket{V}$) is the
polarization state of a single photon: horizontal (vertical) and the
condition
\begin{equation}
\cos \phi=\alpha
\end{equation} is
satisfied.

The polarization degenerate photon pairs generated via type-I
spontaneous parametric down-conversion (SPDC) in two $0.5$mm-thick
nonlinear-$\beta$-barium-borate (BBO) crystals cut at $29.41^o$,
pumped by a $400.8$nm CW diode laser with up to $100$mW of power.
For 1D QWs, triggering on one photon prepares the other photon pairs
at wavelength $801.6$nm into a single-photon state $\ket{\psi_\pm}$
via a polarizing beam splitter (PBS) following by waveplates.
Interference filters determine the photon bandwidth $3$nm and then
pairs of down-converted photons are steered into the different
optical modes (up and down) of the linear-optical network formed by
a series of birefringent calcite beam displacers (BDs) and wave
plates. Output photons are detected using avalanche photo-diodes
(APDs, $7$ns time window) with dark count rate of $<100$s$^{-1}$
whose coincident signals---monitored using a commercially available
counting logic---are used to post-select two single-photon events.
The total coincident counts are about $1000$s$^{-1}$ (the coincident
counts are collected over $40$s). The probability of creating more
than one photon pair is less than $10^{-4}$ and can be neglected.

\begin{figure}
\includegraphics[width=8.5cm]{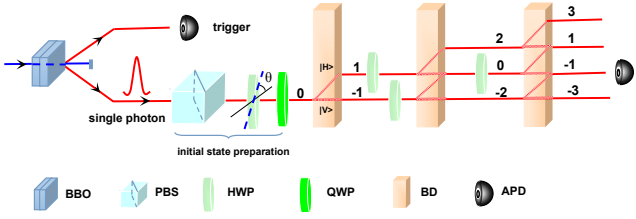}
\caption{(Color online.) Experimental schematic. Detailed sketch of
the setup for realization of unambiguous state discrimination of two
equally probable single-qubit states via a three-step QW .
Single-photons are created via SPDC in two BBO crystals. One photon
in the pair is detected to herald the other photon, which is
injected into the optical network. Arbitrary initial coin states are
prepared by a PBS, HWP and QWP. Position-dependent coin flipping is
realized by the HWP with different setting angle placed in different
optical path. Coincident detection of trigger and heralded photons
at APDs yields data for the QW.}\label{Fig1}
\end{figure}

A standard model of a 1D discrete-time QW consists of a walker
carrying a coin which is flipped before each step. In the basis of
the coin state $\{\ket{H},\ket{V}\}$, the position-dependent coin
flipping operation for the $n$th step is given by
\begin{equation}
C(x,n,\vartheta_{x,n})=\ket{x}\bra{x}\otimes\begin{pmatrix}\cos2\vartheta_{x,n} & \sin2\vartheta_{x,n}\\
        \sin2\vartheta_{x,n} & -\cos2\vartheta_{x,n}\end{pmatrix}
\end{equation}
with $\vartheta_{x,n}\in\left[0,\pi/4\right]$, consisting of a
polarization rotation, which is realized with a half-wave plate
(HWP) placed in the spatial modes $x$ of the interferometer. The
value of $\vartheta$ is determined by the angle between the optic
axis of HWP and horizontal. For some specific evolution, one can
also use identity operator $\one_x=\ket{x}\bra{x}\otimes\one_c$ for
a coin flipping.

The walker's positions are represented by longitudinal spatial
modes. The conditional position shift due to the outcome of the coin
flipping for each step
\begin{equation}
T=\sum_{x}\ket{x+1}\bra{x}\otimes\ket{H}\bra{H}+\ket{x-1}\bra{x}\otimes\ket{V}\bra{V}
\end{equation}
acts on these modes, which is implemented by a BD with length $28$mm
and clear aperture $10$mm$\times 10$mm. The optical axis of each BD
is cut so that vertically polarized photons are directly transmitted
and horizontal photons moves up a $2.7$mm lateral displacement into
a neighboring mode which interferes with the vertical light in the
same mode. Each pair of BDs forms an interferometer. BDs are placed
in sequence and need to have their optical axes mutually aligned.
Co-alignment ensures that beams split by one BD in the sequence
yield maximum interference visibility after passing through a HWP
and the next BD in the sequence. In our experiment, we attain
interference visibility of $0.998$ for each step. The unitary
operation for the $n$th step is $U_n=T\left[\sum_x
C(x,n,\vartheta_{x,n})+\sum_{x'}\one_{x'}\right].$

Back to the unambiguous state discrimination, it can be realized
with a three-step QW. First the initial coin state $\ket{\psi(0)}_c$
is initially prepared in $\ket{\psi_\pm}$ and the walker starts from
the original position $x=0$. For the first step, the coin operation
is identity and the unitary operation is $U_1=T$. After going
through the first BD, the photons are injected into the spatial mode
$\pm1$ and the state of the walker+coin is
\begin{equation}
\ket{\psi_\pm(1)}=\cos\frac{\phi}{2}\ket{1}\ket{H}\pm\sin\frac{\phi}{2}\ket{-1}\ket{V}.
\end{equation} The unitary operation for the second step is
\begin{align}
U_2=&T\Big[\sum_{x\neq\pm1}\ket{x}\bra{x}\otimes\one_c+C(-1,2,\frac{\pi}{4})\nonumber\\
&+C\left(1,2,\frac{1}{2}\arccos{\sqrt{1-\tan^2{\frac{\phi}{2}}}}\right)\Big],
\end{align}
and the state of the walker+coin is
\begin{equation}
\ket{\psi_\pm(2)}=\sqrt{\cos
\phi}\ket{2}\ket{H}+\sin\frac{\phi}{2}\ket{0}(\ket{V}\pm\ket{H}).
\end{equation}
Finally after the transformation for the third step
\begin{equation}
U_3=T\left[C(0,3,\frac{\pi}{8})+\sum_{x\neq0}\ket{x}\bra{x}\otimes\one_c\right]
\end{equation} the finial state is
\begin{align}
&\ket{\psi_+(3})=\sqrt{\cos
\phi}\ket{3}\ket{H}+\sqrt{2}\sin\frac{\phi}{2}\ket{1}\ket{H},\nonumber\\
&\ket{\psi_-(3})=\sqrt{\cos
\phi}\ket{3}\ket{H}-\sqrt{2}\sin\frac{\phi}{2} \ket{-1}\ket{V}
\end{align} During the processing, only the
initial coin state $\ket{\psi_\pm}$ and the coin flipping in the
position $x=1$ for the second step $C(1,2,\vartheta_{1,2})$ depend
on the choice of the states which need to be discriminated, which
decreases the difficulty of the experimental realization.

Corresponding to the three outcomes of the measurement on the
walker's position, which is realized in our experiment by the
coincidence measurement of the photons in the three spatial modes
$x=3,1,-1$ and the trigger photons, we can discriminate the
nonorthogonal states $(\ket{0},\alpha\ket{0}+\beta\ket{1})$. If the
walker is measured in the position $x=\pm1$ we know that the coin
state is $\ket{\psi_\pm}$ which corresponds to the state $\ket{0}$
(or $\alpha\ket{0}+\beta\ket{1}$). If the walker is measured in the
position $x=3$ we know nothing about the state discrimination. The
successful probability theoretically is
\begin{equation}
\eta=2\sin^2\frac{\phi}{2}=1-\alpha,
\end{equation} which increases with $\alpha$ decreasing.
For the extreme case of $\alpha=0$ the measurement becomes a
projective measurement and the probability of discriminating the
states is $1$.

\begin{figure}
\includegraphics[width=8.5cm]{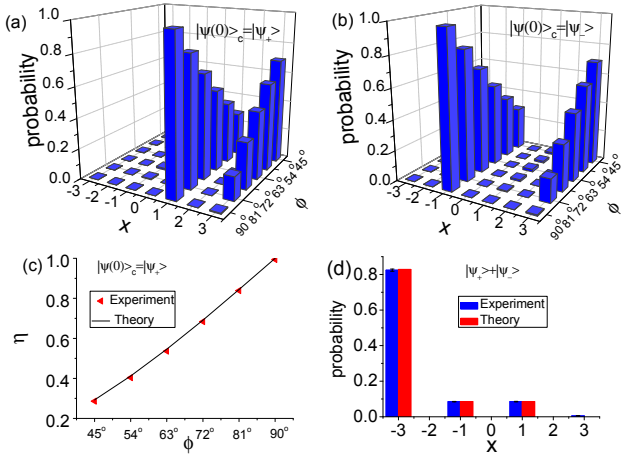}
\caption{(Color online.) Experimental data of the unambiguous state
discrimination via a photonic QW. Measured probability distributions
of the three-step QW with the position-dependent coin and initial
coin state $\ket{\psi_+}$ in (a) and $\ket{\psi_-}$ in (b); the
different coefficients $\phi$ of $\ket{\psi_\pm}$ for unambiguous
state discrimination. (c) Measured successful probability $\eta$
v.s. the coefficient $\phi$ which is related to the state to be
discriminated, compared to theoretical predictions. Error bars are
smaller than used symbols. (d) Probability distribution of the
three-step QW with the initial coin state $\ket{H}$ which is an
equally-weighted superposition of $\ket{\psi_\pm}$ with $\phi=45^o$.
The blue and red bars show the experimental data and theoretical
predictions, respectively.}\label{Fig2}
\end{figure}

The realization of unambiguous state discrimination of two equally
probable single-qubit states via three-step QW are shown in Fig.~1
in detailed. The measured probability distributions for $1$ to $3$
steps of QW are shown in Fig.~2. The probabilities are obtained by
the normalizing coincidence counts on each mode to total for the
respective step. We characterize the quality of the experimental QW
by its 1-norm distance~\cite{B10} from the theoretical predictions
according to
$d=\frac{1}{2}\sum_x\left|P^\text{exp}(x)-P^\text{th}(x)\right|$
shown in Table I. In our experiment the small distances ($d<0.02$)
demonstrate strong agreement between the measured distribution and
theoretic prediction after three steps.

\begin{widetext}
\begin{center}
\begin{table}[htbp]
\begin{tabular}{l|l|l|l|l|l|l|l}
  \hline
  $\alpha$ & $\phi$ & $\ket{\psi(0)}_c$& $\vartheta_{-1,2}$ & $\vartheta_{1,2}$ & $\vartheta_{0,3}$ & $\eta\pm\Delta\eta$ & $d\pm\Delta d$ \\
  \hline\hline
  $0.707$ & $45^o$ & $\ket{\psi_+}$ &  $45^o$ & $12^o14'$ & $22^o30'$ & $0.2861\pm0.0030$ & $0.0171\pm0.0046$ \\ \hline
  $0.588$ & $54^o$ & $\ket{\psi_+}$  & $45^o$ & $15^o19'$ & $22^o30'$ & $0.4037\pm0.0038$ & $0.0184\pm0.0047$ \\ \hline
  $0.454$ & $63^o$ & $\ket{\psi_+}$ & $45^o$ & $18^o54'$ & $22^o30'$ & $0.5362\pm0.0045$ & $0.0192\pm0.0047$ \\ \hline
  $0.309$ & $72^o$ & $\ket{\psi_+}$ & $45^o$  & $23^o18'$ & $22^o30'$ & $0.6834\pm0.0054$ & $0.0127\pm0.0046$\\ \hline
  $0.156$ & $81^o$ & $\ket{\psi_+}$ & $45^o$ & $29^o20'$ & $22^o30'$ & $0.8384\pm0.0062$ & $0.0066\pm0.0044$ \\ \hline
  $0$ & $90^o$ & $\ket{\psi_+}$ & $45^o$ & $45^o$ & $22^o30'$ & $0.9940\pm0.0070$ & $0.0060\pm0.0035$  \\ \hline
  $0.707$ & $45^o$ & $\ket{\psi_-}$  & $45^o$ & $12^o14'$ & $22^o30'$ &$0.2875\pm0.0031$  & $0.0152\pm0.0045$ \\ \hline
  $0.588$ & $54^o$ & $\ket{\psi_-}$  & $45^o$ & $15^o19'$ & $22^o30'$ & $0.4066\pm0.0039$ & $0.0156\pm0.0047$ \\ \hline
  $0.454$ & $63^o$ & $\ket{\psi_-}$ & $45^o$ & $18^o54'$ & $22^o30'$ & $0.5365\pm0.0045$ & $0.0183\pm0.0046$ \\ \hline
  $0.309$ & $72^o$ & $\ket{\psi_-}$ & $45^o$  & $23^o18'$ & $22^o30'$ & $0.6854\pm0.0055$ & $0.0136\pm0.0049$\\ \hline
  $0.156$ & $81^o$ & $\ket{\psi_-}$  & $45^o$ & $29^o20'$ & $22^o30'$ & $0.8394\pm0.0063$ & $0.0071\pm0.0043$ \\ \hline
  $0$ & $90^o$ & $\ket{\psi_-}$ & $45^o$ & $45^o$ & $22^o30'$ & $0.9920\pm0.0071$ & $0.0080\pm0.0036$  \\ \hline
\end{tabular}
\caption{The coefficients of the states to be discriminated, the
initial coin states, the corresponding parameters for the HWP
settings, the experimental data for the successful probabilities and
1-norm distance from the theoretical predictions. Error bars
indicate the statistical uncertainty.}
\end{table}
\end{center}
\end{widetext}

In our experiment, we choose different coefficient $\alpha$ for the
set of two states $\{\ket{0},\alpha\ket{0}+\beta\ket{1}\}$ and
prepare the initial coin state to the corresponding state
$\ket{\psi_\pm}$ with the condition in Eq.~(2) satisfied. For either
of the two states, the photons undergoing through the interferometer
network are measured at the modes $x=3$ for inclusive result and
$x=\pm1$ for conclusive results. Two pronounced peaks for each
$\phi$ shown in the probability distribution in Figs.~2(a) and (b)
clearly prove the demonstration of the unambiguous state
discrimination. With $\alpha$ decreasing from $1/\sqrt{2}$ to $0$
the measured successful probability $\eta$ of the discrimination of
the state $\ket{0}$ increases from  $0.2861\pm0.0030$ to
$0.9940\pm0.0070$ (from $0.2875\pm0.0031$ to $0.9920\pm0.0071$ to
discriminate the state $\alpha\ket{0}+\beta\ket{1}$) shown in
Fig.~2(c). The process can be deterministic if and only if the two
states are orthogonal, i.e. $\alpha=0$, and the measurement becomes
the projective measurement.

Taking a superposition of $\ket{\psi_\pm}$ as an initial coin state
$\ket{\psi(0)}_c\propto a\ket{\psi_+}+b\ket{\psi_-}$
($a,b\in\mathbb{R}$) the photons undergoing through the
interferometer network for the three-step QW are detected in the
positions $x=1$ and $x=-1$ simultaneously with probabilities
$2a^2\sin^2\frac{\phi}{2}/(a^2+b^2+2ab\cos\phi)$ and
$2b^2\sin^2\frac{\phi}{2}/(a^2+b^2+2ab\cos\phi)$ respectively for
the conclusive results and in the position $x=3$ with the
probability $4\cos\phi/(a^2+b^2+2ab\cos\phi)$ for the inconclusive
result. The ratio of the probabilities for the two conclusive
results is $a^2/b^2$, which is also proven in our experiment. In
Fig.~2(d) we show with $a=b$ the probabilities of $x=1$ and $x=-1$
are measured approximately equal, i.e. $P(1)=0.0854\pm0.0015$ and
$P(-1)=0.0850\pm0.0015$.

Therefore we have clearly demonstrated the generalized measurement
scheme via a three-step QW for the first time. The unambiguous state
discrimination is confirmed by direct measurement and found to be
consistent with the ideal theoretical values at the level of the
average distance $d<0.02$ and the fidelity of the coin state
measured in the position $x=\pm1$ $F>0.9911$.

Now based on our experimental realization and result, we go for a QW
algorithm for generation of the POVM elements corresponding to the
unambiguous state discrimination. Let us consider the POVM element
$E_i (i=\pm1)$.

\begin{enumerate}

\item Start with the state $\ket{i}\ket{\psi_i}_c$.

\item Apply the reversed evolution operator of the three-step QW
$\tilde{U}^\dagger_1 \tilde{U}^\dagger_2 \tilde{U}^\dagger_3$ on
$\ket{i}\ket{\psi_i}_c$ with
$\tilde{U}^\dagger_n=C^\dagger(x,n,\vartheta_{x,n})T^\dagger$.

\item Take the overlap with the ancilla state (the initial state of the
walker) $\ket{x=0}$ and obtain
$\ket{\tilde{\psi}_i}=\bra{0}\tilde{U}^\dagger_1 \tilde{U}^\dagger_2
\tilde{U}^\dagger_3\ket{i}\ket{\psi_i}_c\bra{\psi_i}\bra{i}\tilde{U}_3
\tilde{U}_2 \tilde{U}_1\ket{0}$.

\item The POVM element for unambiguous state discrimination is then
$E_i=\ket{\tilde{\psi_i}}\bra{\tilde{\psi_i}}$.

\end{enumerate}
To prove that, we have \begin{widetext}
\begin{align}
&E_+\ket{\psi_-}=\frac{1}{2\cos^2\frac{\phi}{2}}(\sin\frac{\phi}{2}\ket{H}+\cos\frac{\phi}{2}\ket{V})(\sin\frac{\phi}{2}\bra{H}+\cos\frac{\phi}{2}\bra{V})\ket{\psi_-}=0\nonumber\\
&E_-\ket{\psi_+}=\frac{1}{2\cos^2\frac{\phi}{2}}(-\sin\frac{\phi}{2}\ket{H}+\cos\frac{\phi}{2}\ket{V})(-\sin\frac{\phi}{2}\bra{H}+\cos\frac{\phi}{2}\bra{V})\ket{\psi_+}=0.
\end{align}
\end{widetext}

We experimentally prove that discrete-time QWs are capable of
performing generalized measurements on a single qubit. We explicitly
for the first time realize a photonic three-step QW with
position-dependent coin flipping for unambiguous state
discrimination of two equally probable single-qubit states with
single photons undergoing through an interferometer network.
Furthermore, corresponding to the unambiguous state discrimination
problem the QW generation of the POVM elements is shown.

We use a very simple experiment to show the meaning of the idea on
using QW to implement a generalized measurement which is shown
in~\cite{KW13}. In a QW, the position shifts of the walker depend on
the coin state being measured. That is, a QW is based on projection
measurement which in the QW scheme can be extended in the evolution
time. Furthermore as shown in the above experiment, the projection
measurement model can naturally be extend to an arbitrary
generalized measurement without the extension of the Hilbert space.

\acknowledgements We would like to thank C.F. Li and Y.S. Zhang for
stimulating discussions. This work has been supported by NSFC under
11174052 and 11474049, the Open Fund from the SKLPS of ECNU and 973
Program under 2011CB921203.

\end{document}